\pdfoutput=1
\documentclass[letterpaper,journal]{IEEEtran}
\usepackage{graphicx}
\usepackage{amsmath,amssymb}
\usepackage{booktabs}
\usepackage{stfloats}
\usepackage{array}
\usepackage{cite}
\usepackage[table]{xcolor}

\begin{document}

\title{Reinsertion-Aware Parameter Estimation of the WECC Composite Load Model via Closed-Loop Equivalent-Network Co-Estimation}

\author{Jong Ha Woo,
        Victor Daldegan Paduani,
        Zishuo Yang,
        Yu Ma,
        Qi Xiao,
        and Ning Lu
\thanks{J. H. Woo, Z. Yang, Y. Ma, Q. Xiao, and N. Lu are with the Department of Electrical and Computer Engineering, North Carolina State University, Raleigh, NC 27606, USA (e-mail: jwoo6@ncsu.edu; zyang66@ncsu.edu; yma47@ncsu.edu; qxiao3@ncsu.edu; nlu2@ncsu.edu).}%
\thanks{V. D. Paduani is with the New York Power Authority, White Plains, NY 10603, USA (e-mail: Victor.DaldeganPaduani@nypa.gov).}%
\thanks{This work has been submitted to the IEEE for possible publication. Copyright may be transferred without notice, after which this version may no longer be accessible.}}

\markboth{Preprint -- submitted to IEEE Transactions on Power Systems}{Woo \MakeLowercase{\textit{et al.}}: Reinsertion-Aware Parameter Estimation of the WECC CMLD}

\maketitle

\begin{abstract}
Measurement-based parameter estimation of the WECC composite load model (CMLD) commonly relies on voltage replay, where measured voltage is imposed on an isolated CMLD and its parameters are optimized to reproduce measured active and reactive power. However, this open-loop formulation neglects the two-way dynamic interaction between the load and grid, potentially causing large errors when the estimated model is reinserted into dynamic grid simulations. To address this limitation, we propose a closed-loop parameter estimation framework that alternately estimates the CMLD parameters and a five-parameter equivalent network representing the external grid. This framework restores load-grid interaction during estimation and enables voltage, active power, and reactive power to jointly constrain the parameter estimation. Compared with four baseline methods, the proposed approach achieves the highest reinsertion accuracy in both the CMLD benchmark and transmission and distribution (T\&D) co-simulation environments. In the T\&D case, it reduces active-power, reactive-power, and voltage errors by 73\%, 64\%, and 62\%, respectively. The identified model also maintains improved reinsertion accuracy under disturbances not used for estimation, demonstrating transferability beyond the identification event.

\end{abstract}


\section{Introduction}
\IEEEPARstart{L}{oad} dynamics play an important role in power system stability. A well-known example is fault-induced delayed voltage recovery (FIDVR), where stalled air-conditioner motors draw large reactive power and delay voltage recovery after a fault \cite{williams}. To represent such dynamic load behavior, the WECC composite load model (CMLD) combines static loads, power-electronic loads, and multiple types of induction motors behind an equivalent feeder and transformer \cite{wecc}. It is the standard dynamic load model used by WECC planners and utilities in transient-stability and FIDVR studies, where its parameters directly shape the simulated voltage recovery. However, the CMLD contains more than 130 parameters, many of which cannot be directly measured in the field. Accurate parameter estimation is therefore essential for using the CMLD in dynamic simulation.

Measurement-based methods typically estimate CMLD parameters through \emph{voltage replay} \cite{regulski}. A recorded voltage trajectory is applied to an isolated CMLD, and its parameters are adjusted until the simulated active and reactive power match the measurements. However, a CMLD is ultimately intended to be used within a power system simulation. Once the estimated model is reinserted into the grid, its active and reactive power affect the system voltage, which in turn affects the load response. We refer to this closed-loop use of the estimated model as \emph{reinsertion}. Therefore, a CMLD that closely matches measured P and Q under voltage replay may not necessarily reproduce the same dynamic response after reinsertion.

Table~\ref{tab:prior} compares existing CMLD parameter-estimation methods from three perspectives important to practical application: \emph{data and model requirements}, \emph{estimation formulation}, and \emph{reinsertion validation}. First, estimation from a single disturbance is desirable because severe events are infrequent and load composition may vary with operating conditions and time \cite{ambient}. Second, requiring the full dynamic network model limits practical applicability: an accurate model may not be available at the measurement location, and repeatedly simulating a large network during parameter search can be computationally expensive. These considerations favor methods that can estimate CMLD parameters from a single event without requiring the full network model.

Existing methods address these practical requirements to different degrees. Supervised learning requires labeled parameter sets \cite{resdnn}; ambient-data and diffusion-based methods use measurements spanning multiple operating conditions or events \cite{ambient,diffusion}; and estimation can also be performed directly within a full network model \cite{ddqn}. Most measurement-based methods, however, use a common replay-based formulation, despite employing different optimization, reinforcement-learning, Bayesian, or machine-learning techniques \cite{regulski,kim,edrl,transferq,seqbayes,amortized}. In voltage replay, the measured voltage is imposed as a fixed input and the CMLD parameters are estimated by matching the measured $P$ and $Q$. Because voltage is prescribed rather than reproduced, the load-grid interaction is absent from the estimation process and voltage cannot provide an independent fitting objective. Consequently, a small replay error does not necessarily imply that the estimated CMLD will reproduce the measured system response after reinsertion.

\begin{table*}[!t]
\centering
\caption{Comparison of CMLD Parameter-Estimation Methods}
\label{tab:prior}
\footnotesize
\setlength{\tabcolsep}{4.5pt}
\begin{tabular}{lccccccc}
\toprule
\textbf{Ref.} &\textbf{ Method} & \textbf{Single Event} & \textbf{Requires Full Network} & \textbf{Min $P,Q$ error} & \textbf{Min $V$ error} & \textbf{Closed Loop }& \textbf{Reinsertion Validation} \\
\midrule
\cite{resdnn} & ML & Y & N & N & N & N & N \\
\cite{ambient} & Optimization & N & N & Y & N & N & N \\
\cite{diffusion} & ML & N & N & Y & N & N & N \\
\cite{ddqn} & RL & Y & Y & Y & N & n/a & n/a \\
\cite{regulski,kim} & Optimization & Y & N & Y & N & N & N \\
\cite{edrl,transferq} & RL & Y & N & Y & N & N & N \\
\cite{seqbayes} & Bayesian & Y & N & Y & N & N & N \\
\cite{amortized} & ML & Y & N & Y & N & N & N \\
\midrule
\rowcolor{lightgray}
\textbf{This paper} & \textbf{Co-estimation} & \textbf{Y} & \textbf{N} & \textbf{Y} & \textbf{Y} & \textbf{Y} & \textbf{Y} \\
\bottomrule
\end{tabular}

\vspace{2pt}
\end{table*}

This distinction motivates the last three columns of Table~\ref{tab:prior}. For an estimated CMLD to reproduce system-level dynamics, its $P$ and $Q$ response must interact with the grid voltage, and the resulting $P$, $Q$, and $V$ trajectories should be evaluated after the model is reinserted into a dynamic simulation. As summarized in Table~\ref{tab:prior}, the measurement-based methods considered here do not jointly provide closed-loop estimation and reinsertion validation without requiring the full network model. The resulting gap between replay accuracy and reinsertion accuracy therefore remains an important issue for measurement-based CMLD parameter estimation.

To address this gap, this paper replaces the external grid behind the measurement point with a five-parameter equivalent network and alternately co-estimates the equivalent-network and CMLD parameters. This creates a closed loop in which the estimated load response affects the voltage and the resulting voltage, in turn, affects the load response. Voltage therefore becomes an output of the estimation process rather than a prescribed input, allowing the $P$, $Q$, and $V$ errors to be jointly minimized. Because the external grid is represented by a two-bus equivalent, the proposed approach captures the load-grid interaction without requiring repeated full-network dynamic simulations.

The main contributions of this paper are summarized as follows.
\begin{enumerate}

\item We propose a closed-loop CMLD parameter-estimation framework that alternately estimates the CMLD parameters and a five-parameter equivalent network. The equivalent network restores the dynamic interaction between the load and external grid without requiring the full network model, making voltage a model-generated output rather than a prescribed input. This formulation enables $P$, $Q$, and $V$ to jointly constrain the estimated CMLD and directly targets the dynamic response relevant to reinsertion.

\item We quantify the gap between \emph{replay accuracy} and \emph{reinsertion accuracy} in measurement-based CMLD parameter estimation. Results show that parameter sets with nearly indistinguishable replay responses can exhibit substantial differences in both parameter values and closed-loop responses after reinsertion. This demonstrates that a small replay error alone does not guarantee either accurate parameter recovery or accurate system-level dynamic response.

\item We validate the proposed framework using both a controlled CMLD benchmark with known ground-truth parameters and a transmission and distribution (T\&D) co-simulation in which a detailed distribution feeder serves as the ground truth. In the T\&D case, the proposed method reduces the $P$, $Q$, and $V$ reinsertion errors by 73\%, 64\%, and 62\%, respectively, for the identification event, while maintaining improved average performance across unseen fault events. These results demonstrate improved transferability beyond the disturbance used for parameter estimation.

\end{enumerate}

The remainder of this paper is organized as follows. Section~II formulates the CMLD parameter-estimation problem, analyzes the limitations of voltage replay, and introduces the proposed closed-loop co-estimation method. Section~III presents the validation setup and results, and Section~IV concludes the paper.

\section{Methodology}
\subsection{CMLD Parameter Estimation}
\begin{figure}[!h]
\centering
\includegraphics[width=\columnwidth]{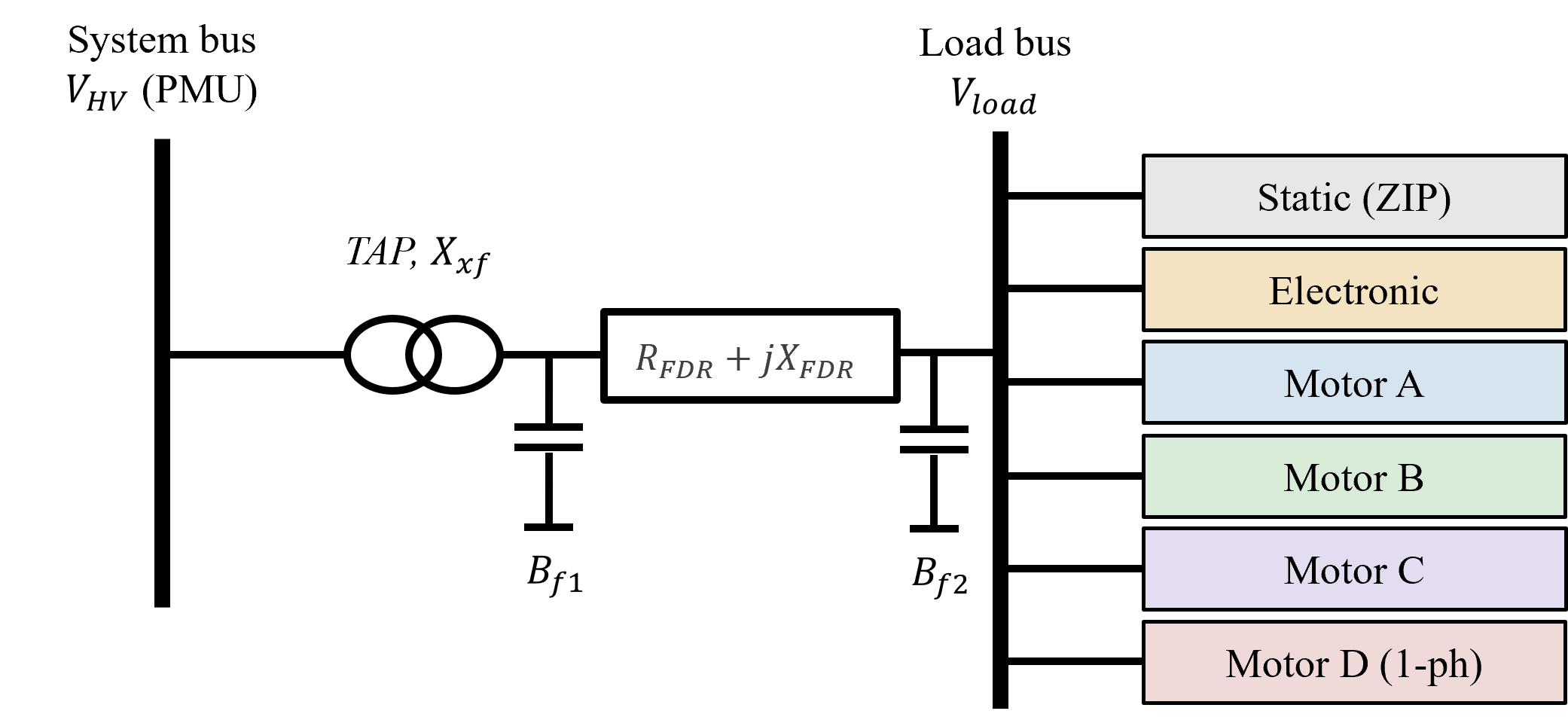}
\caption{One-line structure of the WECC composite load model: substation transformer, $\pi$-equivalent feeder, and six load components.}
\label{fig:structure}
\end{figure}
\begin{figure*}[!tb]
\centering
\includegraphics[width=\textwidth]{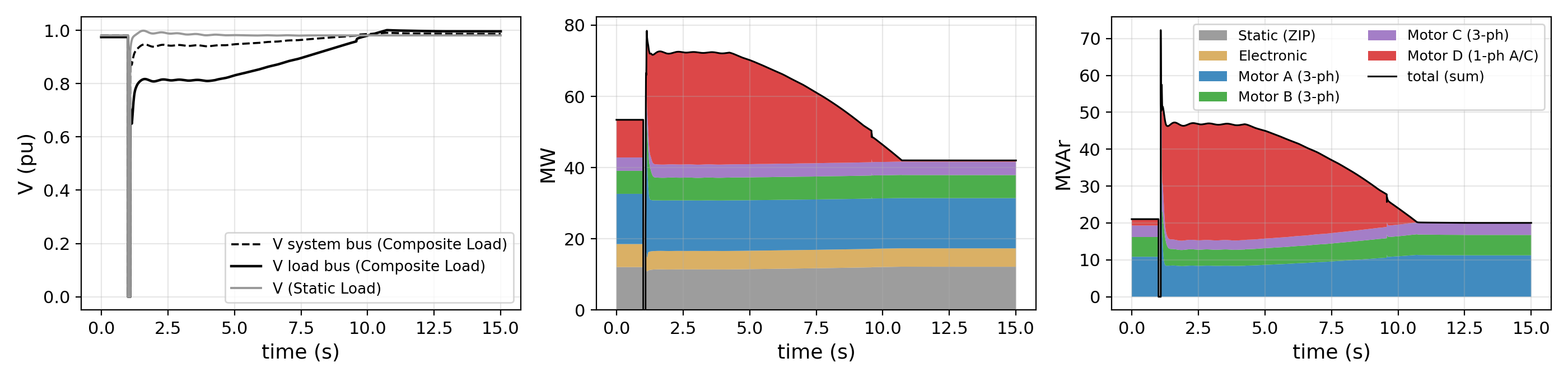}
\caption{CMLD fault response: bus voltages versus a static load (left) and per-component P, Q decomposition (center, right).}
\label{fig:fidvr}
\end{figure*}
The WECC composite load model (CMLD), illustrated in Fig.~\ref{fig:structure}, represents the aggregated load behind a transmission-level measurement point using a substation transformer, an equivalent feeder, and parallel static, power-electronic, three-phase motor, and single-phase motor components \cite{wecc}. The detailed component models and their dynamic equations follow the WECC specification and are not repeated here.

Let $\theta$ denote the CMLD parameters to be estimated from the measured voltage and power trajectories
\begin{equation}
\mathbf{z}^{m}(t)=
[V^{m}(t),P^{m}(t),Q^{m}(t)].
\end{equation}
For a given $\theta$, the aggregate load response is
\begin{equation}
P_L(t;\theta)=\sum_k P_k(t;\theta),\qquad
Q_L(t;\theta)=\sum_k Q_k(t;\theta),
\label{eq:cmld}
\end{equation}
where $k$ indexes the CMLD components.

As estimating all CMLD parameters from a single disturbance is generally impractical, this study follows prior work \cite{ddqn,amortized} and estimates a 50-dimensional subset containing the influential component dynamics, load-composition fractions, and feeder parameters. The feasible parameter space is
\begin{equation}
\Theta=\{\theta: \mathbf{lb}\leq\theta\leq\mathbf{ub}\}
\cap\Theta_{\mathrm{phy}},
\label{eq:theta}
\end{equation}
where $\Theta_{\mathrm{phy}}$ enforces the physical constraints of the CMLD. Candidate parameters violating these constraints are projected onto the feasible set before simulation.

The component-level response in Fig.~\ref{fig:fidvr} further illustrates an important consideration for CMLD parameter estimation. During a fault-induced delayed voltage recovery (FIDVR) event, undervoltage causes the single-phase motor to stall and draw substantial reactive power, but this contribution diminishes as thermal protection progressively trips the stalled motors. In contrast, the response of power-electronic resources, such as distributed PV, may persist throughout the post-fault period as determined by their low-voltage ride-through and recovery characteristics. Consequently, similar aggregate $P$ and $Q$ deviations immediately after a fault may originate from components with substantially different temporal behaviors. Accurate parameter estimation should therefore reproduce not only the magnitude of the aggregate response but also its post-fault evolution, so that transient motor-stall dynamics can be distinguished from the more persistent response of power-electronic resources. This distinction is particularly important for reproducing the depth and duration of FIDVR and the subsequent voltage recovery.
\subsection{Conventional Voltage-Replay Estimation}

Most measurement-based CMLD parameter-estimation methods follow a voltage-replay framework. For a recorded disturbance, the measured boundary-bus voltage $V_m(t)$ is imposed on an isolated CMLD, while the measured active and reactive powers, $P_m(t)$ and $Q_m(t)$, are used as fitting targets. For a candidate parameter vector $\theta$, the replayed response is
\begin{equation}
(\hat P,\hat Q)(t)=M(V_m(t);\theta),
\label{eq:replaymap}
\end{equation}
where $M(\cdot;\theta)$ represents the CMLD dynamics. The conventional replay objective is therefore
\begin{equation}
J_1(\theta)=E_P+E_Q,
\label{eq:j1}
\end{equation}
where $E_P$ and $E_Q$ denote the active- and reactive-power trajectory errors, respectively.

The fundamental characteristic of voltage replay is that the voltage is a prescribed input independent of the candidate CMLD parameters:
\begin{equation}
V(t)=V_m(t), \qquad \forall\theta\in\Theta.
\label{eq:fixedV}
\end{equation}
Changing $\theta$ can therefore change $\hat P$ and $\hat Q$, but cannot change the voltage applied to the CMLD. As a result, the feedback path from the load response to the grid voltage is absent during estimation, and the measured voltage provides no independent information for distinguishing candidate CMLD parameters. A parameter set can thus achieve a small $P$-$Q$ replay error without necessarily reproducing the measured voltage or the coupled $P$-$Q$-$V$ response after reinsertion.

\subsection{Proposed Closed-Loop Equivalent-Network Co-Estimation}

\subsubsection{From Voltage Replay to Closed-Loop Estimation}

The key difference between the proposed method and conventional voltage replay is how the grid-side voltage is represented during parameter estimation. As discussed in Section~II-B, conventional methods directly impose the measured voltage trajectory $V_m(t)$ on the candidate CMLD. The grid is therefore absent from the estimation loop, and the voltage applied to the CMLD remains unchanged regardless of the candidate parameters $\theta$. Consequently, the effect of the estimated load on the system voltage cannot be captured during parameter estimation.

\begin{figure}[!h]
\centering
\includegraphics[width=\columnwidth]{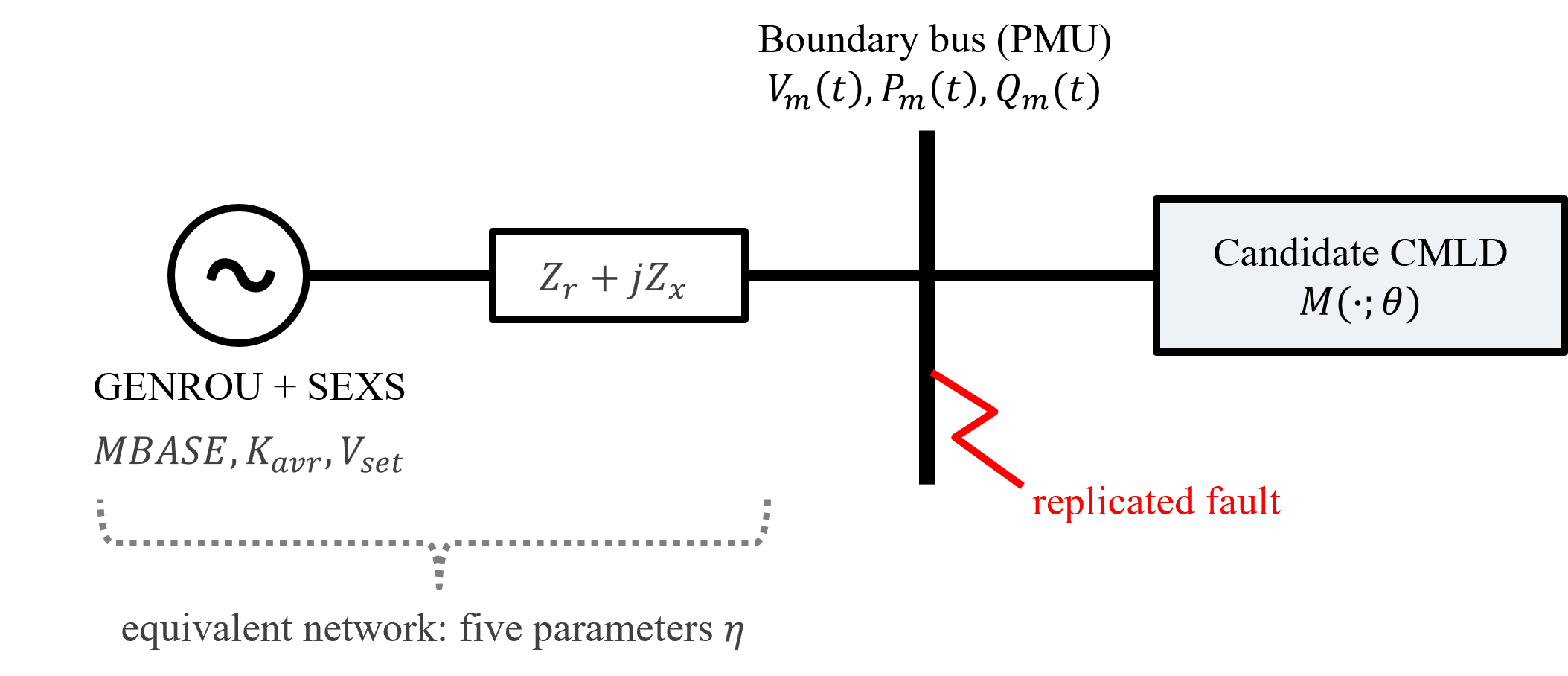}
\caption{Five-parameter equivalent network: the closed loop that makes the voltage an output.}
\label{fig:eqnet}
\end{figure}

The proposed method replaces this prescribed voltage trajectory with a low-order equivalent network representing the external grid, as illustrated in Fig.~\ref{fig:eqnet}. Instead of directly imposing $V_m(t)$ on the CMLD, the measured disturbance is reproduced in the equivalent system and the boundary-bus voltage is solved dynamically. The candidate CMLD therefore interacts with the equivalent network: a change in $\theta$ changes the load $P$ and $Q$, which changes the boundary voltage, and the resulting voltage in turn changes the subsequent load response. The estimation process is thus converted from open-loop voltage replay to a closed-loop load-grid simulation.

This closed-loop formulation provides two important advantages. First, the estimated CMLD is evaluated under the same type of load-grid interaction that occurs after reinsertion into a power system simulation. Second, because voltage is now a model output rather than a prescribed input, its deviation from the measured voltage can be included in the estimation objective together with the $P$ and $Q$ errors. The resulting parameters are therefore identified to reproduce the coupled $P$-$Q$-$V$ dynamic response, rather than only the $P$ and $Q$ response under a fixed voltage trajectory.

\subsubsection{Equivalent-Network Model}

To restore this interaction without requiring the full network model, the external grid behind the measurement point is represented by a low-order equivalent consisting of one equivalent source and one equivalent line, as shown in Fig.~\ref{fig:eqnet}. The source is represented by the PSS/E GENROU synchronous-machine model with a simplified SEXS excitation system, and the equivalent line is represented by $Z_r+jZ_x$. The candidate CMLD is connected at the boundary bus where the measurements are available.

The resulting closed-loop dynamic response can be expressed as
\begin{equation}
(\hat V,\hat P,\hat Q)(t)=S(\theta,\eta),
\label{eq:eqnet1}
\end{equation}
where $\theta$ denotes the CMLD parameter vector and
$\eta=[\mathrm{MBASE},K_{avr},Z_r,Z_x,V_{set}]$ denotes the five-parameter equivalent-network vector. Here, $\mathrm{MBASE}$ characterizes the strength of the equivalent source; $Z_r$ and $Z_x$ represent the resistance and reactance of the equivalent transfer impedance, respectively; and $K_{avr}$ and $V_{set}$ characterize the voltage-regulation response of the equivalent source. The remaining generator and exciter parameters are fixed at representative values \cite{xu}.

The equivalent network is intentionally kept low order. Its purpose is not to reconstruct the external grid in detail, but to reproduce the dominant voltage response required to restore the load-grid feedback during CMLD parameter estimation. Limiting the equivalent to five adjustable parameters also reduces the risk that excessive network-model flexibility compensates for errors in the CMLD parameters.

\subsubsection{Voltage-Inclusive Objective and Co-Estimation}

The closed-loop formulation fundamentally changes the information available for CMLD parameter estimation. In conventional voltage replay, $V_m(t)$ is prescribed for every candidate $\theta$ and therefore carries no independent fitting error. In the proposed formulation, however, the simulated voltage $\hat V$ in \eqref{eq:eqnet1} depends on both the CMLD parameters $\theta$ and the equivalent-network parameters $\eta$. A change in $\theta$ changes the load response, which changes $\hat V$ through the equivalent network; the changed voltage then feeds back into the subsequent CMLD response. The voltage trajectory therefore provides an additional constraint on the candidate CMLD parameters.

Accordingly, the conventional replay objective (\ref{eq:j1})
is extended to the closed-loop objective
\begin{equation}
J_2(\theta,\eta) = E_P+E_Q+w_VE_V,
\label{eq:j2}
\end{equation}
where $E_V$ measures the mismatch between the simulated and measured voltage trajectories and $w_V$ balances the voltage and power error terms. In this work $w_V=100$ ($E_P$, $E_Q$ in MW/MVAr, $E_V$ in p.u.), so a 0.01-p.u. voltage error weighs as much as a 1-MW power error; the same value is used in both environments. Unlike simply adding another fitting metric, $E_V$ is informative here because $\hat V$ is generated by the candidate CMLD interacting with the equivalent network rather than imposed from the measurement. Thus, candidate parameter sets that produce similar $P$ and $Q$ responses under voltage replay can be further distinguished by whether they reproduce the measured closed-loop voltage response.

\begin{figure*}[!t]
\centering
\includegraphics[width=0.95\textwidth]{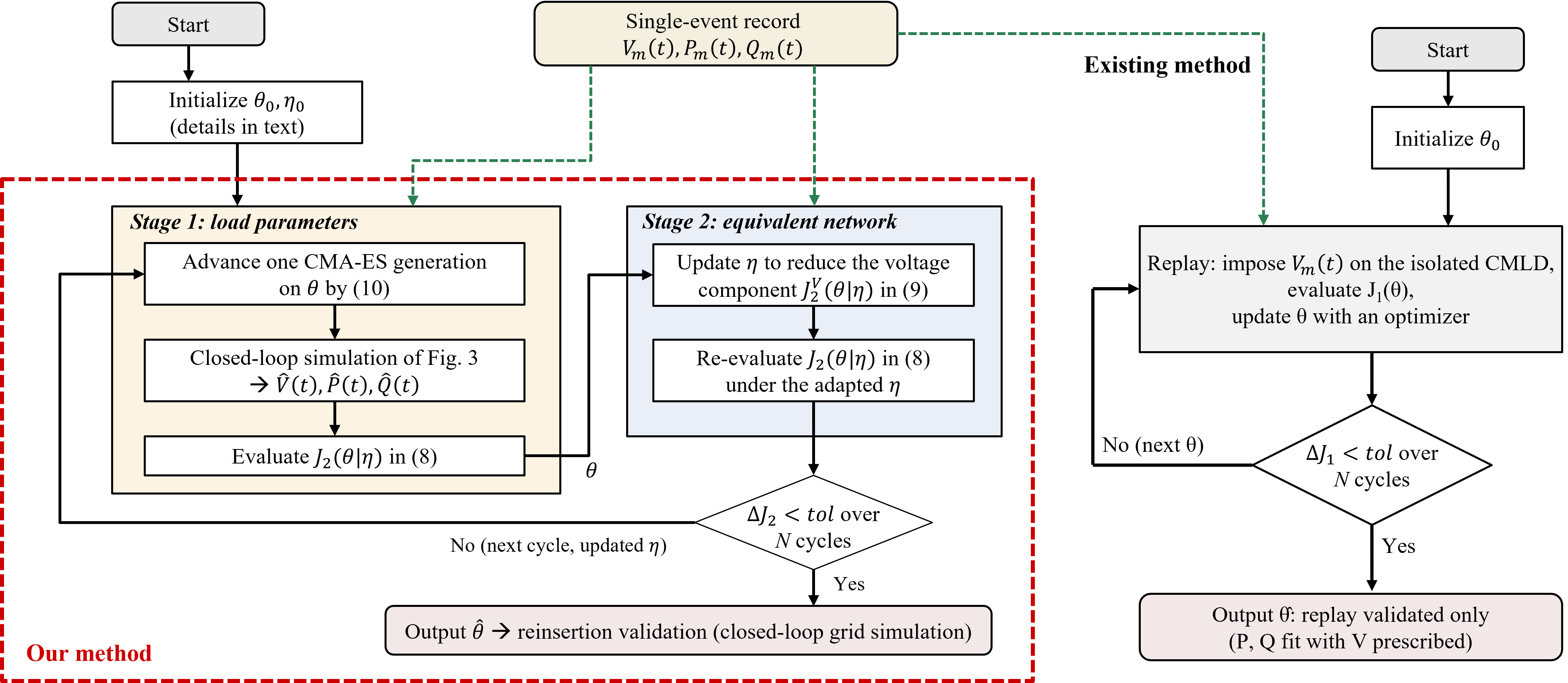}
\caption{Two-stage alternating co-estimation workflow. Stage 1 searches the load parameters with $J_2$ in \eqref{eq:j2}; Stage 2 calibrates the equivalent network with its voltage component $J_2^V$ in \eqref{eq:j2v}.}
\label{fig:workflow}
\end{figure*}

The CMLD and equivalent-network parameters are alternately updated. For a fixed $\eta$, $\theta$ is optimized using the full closed-loop objective $J_2$. For a fixed $\theta$, the equivalent-network parameters are calibrated using
\begin{equation}
J_2^V(\eta\mid\theta)=E_V.
\label{eq:j2v}
\end{equation}
These two steps are repeated until convergence, as illustrated in Fig.~\ref{fig:workflow}. As the equivalent network better reproduces the grid-side voltage response, the voltage term provides a more informative constraint on $\theta$; conversely, as the CMLD estimate improves, the equivalent network can be calibrated with less contamination from load-model error. The alternating procedure therefore jointly refines the two sides of the closed-loop load-grid interaction.

\subsubsection{CMLD Parameter Update}

For a fixed set of equivalent-network parameters $\eta^{(k)}$, the CMLD parameters are updated by solving
\begin{equation}
\theta^{(k+1)}=\arg\min_{\theta\in\Theta}
J_2\left(\theta,\eta^{(k)}\right),
\label{eq:theta_update}
\end{equation}
where $\Theta$ denotes the feasible CMLD parameter space defined in \eqref{eq:theta}. Candidate parameters that violate the physical constraints are projected onto the feasible set before evaluation.

The optimization in \eqref{eq:theta_update} is independent of a particular search algorithm. Any optimizer capable of handling the nonlinear, nonconvex CMLD parameter-estimation problem can in principle be used within the proposed closed-loop framework.

\subsection{Evaluation Metrics and Baseline Methods}

The dynamic-response accuracy of an estimated CMLD is evaluated using the root-mean-square error (RMSE). For $X\in\{P,Q,V\}$,
\begin{equation}
E_X=
\left[
\frac{1}{N}
\sum_{t=1}^{N}
\left(\hat X(t)-X_m(t)\right)^2
\right]^{1/2},
\label{eq:rms}
\end{equation}
where $X_m(t)$ and $\hat X(t)$ denote the measured and simulated trajectories, respectively, and $N$ is the number of samples included in the evaluation. Samples immediately surrounding fault application and clearing are excluded to avoid numerical switching transients.

Two types of dynamic-response errors are distinguished. The \emph{replay error} is evaluated with the measured voltage $V_m(t)$ prescribed at the CMLD boundary. Because voltage is imposed rather than simulated, the replay error contains only the resulting active- and reactive-power errors:
\begin{equation}
\mathbf e_{\mathrm{rep}}(\hat\theta) =
\left[E_P^{\mathrm{rep}},E_Q^{\mathrm{rep}}\right].
\label{eq:ereplay}
\end{equation}
In contrast, the \emph{reinsertion error} is evaluated after the estimated CMLD is reinserted into the original grid model of each validation environment and the disturbance is reproduced. Because voltage is then determined by the closed-loop load-grid interaction, reinsertion accuracy is evaluated using all three trajectories:
\begin{equation}
\mathbf e_{\mathrm{reins}}(\hat\theta) =
\left[E_P^{\mathrm{reins}},E_Q^{\mathrm{reins}},E_V^{\mathrm{reins}}\right].
\label{eq:ereins}
\end{equation}
The equivalent network introduced by the proposed method is used only during parameter estimation and is not used for reinsertion validation.

When the ground-truth CMLD parameters are available, parameter-estimation accuracy is additionally quantified using the span-normalized mean absolute error (NMAE):
\begin{equation}
\mathrm{NMAE}=
\frac{1}{n}
\sum_{i=1}^{n}
\frac{
\left|\hat{\theta}_i-\theta_i^{\mathrm{GT}}\right|
}{
ub_i-lb_i
}
\times100
\label{eq:error}
\end{equation}
where $\theta_i^{\mathrm{GT}}$ and $\hat{\theta}_i$ denote the ground-truth and estimated values of parameter $i$, respectively, and $ub_i$ and $lb_i$ denote its upper and lower search bounds. Because ground-truth CMLD parameters are available only in the CMLD Benchmark case, NMAE is evaluated only for that environment.

Four conventional voltage-replay estimators are implemented as baselines: particle swarm optimization (PSO) \cite{kennedy}, genetic algorithm (GA) \cite{goldberg}, covariance matrix adaptation evolution strategy (CMA-ES) \cite{hansen}, and trust-region-reflective nonlinear least squares (TRF) \cite{coleman}. These methods employ different parameter-search strategies while sharing the same voltage-replay formulation, parameter space, physical constraints, and measurement data. CMA-ES is also used to solve the CMLD parameter-update problem in \eqref{eq:theta_update} within the proposed closed-loop framework. Using the same optimizer for both the conventional CMA-ES baseline and the proposed method provides a controlled comparison in which the principal difference is the estimation formulation rather than the underlying parameter-search algorithm. The implementation settings for all methods are summarized in Table~\ref{tab:protocol}.

\begin{table}[!t]
\centering
\caption{Implementation Settings of the Compared Methods}
\label{tab:protocol}
\footnotesize
\renewcommand{\arraystretch}{1.25}
\setlength{\tabcolsep}{3pt}
\begin{tabular}{lll}
\toprule

\textbf{Method} & \textbf{Search Configuration} & \textbf{Objective} \\
\midrule
PSO \cite{kennedy}
& $N_p=30$, $w=0.72$, $c_1=c_2=1.49$
& $J_1$ \\

GA \cite{goldberg}
& $N_p=40$, tournament-3, BLX-0.5
& $J_1$ \\

CMA \cite{hansen}
& $N_p=20$, $\sigma_0=0.25$
& $J_1$ \\

TRF \cite{coleman}
& Bounded NLS, finite-difference Jacobian
& $J_1$ \\

\rowcolor{lightgray}
Proposed
& \begin{tabular}[t]{@{}l@{}}CMA ($N_p=20$), load parameters $\theta$\\CMA ($N_p=5$), network $\eta$, alternating\end{tabular}
& \begin{tabular}[t]{@{}l@{}}$J_2$\\$J_2^V$\end{tabular} \\
\bottomrule
\end{tabular}
\end{table}

\section{Simulation Results}

\begin{figure}[!h]
\centering
\includegraphics[width=\columnwidth]{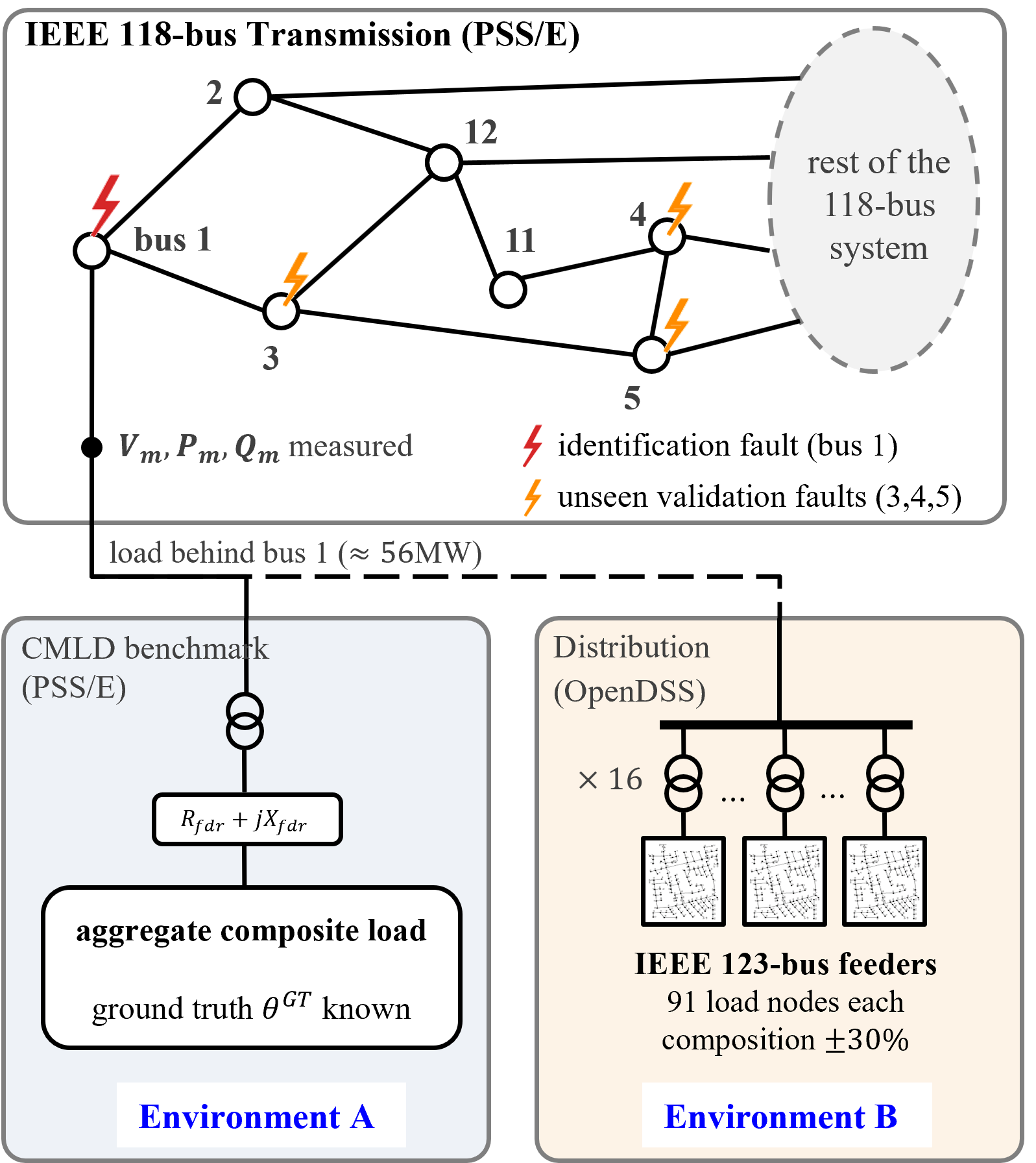}
\caption{Layout of the two validation environments and the fault locations. Both environments share the IEEE 118-bus system and the four fault buses}
\label{fig:envlayout}
\end{figure}

The proposed method is evaluated using two validation environments with different levels of model fidelity. Environment A provides a controlled benchmark in which the ground-truth CMLD parameters are known, enabling direct evaluation of parameter-recovery accuracy. Environment B replaces the CMLD ground truth with a detailed distribution feeder in a T\&D co-simulation, providing a more realistic assessment in which no exact CMLD parameter vector exists. The layout of both environments and the fault locations are shown in Fig.~\ref{fig:envlayout}.

\subsection{Validation Environments}

\begin{table*}[!b]
\centering
\caption{Reinsertion Comparison in Environment A: Errors at Bus 1 for Each Fault Bus ($E_P^{\mathrm{reins}}$ MW / $E_Q^{\mathrm{reins}}$ MVAr / $E_V^{\mathrm{reins}}$ p.u.)}
\label{tab:reinsA}
\footnotesize
\begin{tabular}{lcccc}
\toprule
\textbf{Method} & \textbf{Fault-at-Bus1 (Identification)} & \textbf{Fault-at-Bus3 (unseen event)} & \textbf{Fault-at-Bus4 (unseen event)} & \textbf{Fault-at-Bus5 (unseen event)} \\
\midrule
\rowcolor{lightgray}
Proposed & 0.48 / 0.63 / 0.0006 & 0.49 / 0.65 / 0.0007 & 0.54 / 0.67 / 0.0008 & 0.51 / 0.67 / 0.0008 \\
PSO & 0.93 / 0.84 / 0.0010 & 0.93 / 0.86 / 0.0010 & 0.93 / 0.83 / 0.0010 & 0.94 / 0.83 / 0.0010 \\
GA & 1.97 / 5.60 / 0.0049 & 1.95 / 5.58 / 0.0049 & 1.98 / 5.72 / 0.0069 & 3.09 / 6.05 / 0.0053 \\
CMA & 0.86 / 2.77 / 0.0024 & 0.86 / 2.79 / 0.0025 & 0.86 / 2.74 / 0.0024 & 0.86 / 2.75 / 0.0025 \\
TRF & 0.49 / 1.23 / 0.0010 & 0.50 / 1.27 / 0.0011 & 0.44 / 1.32 / 0.0012 & 0.49 / 1.26 / 0.0011 \\
\bottomrule
\end{tabular}
\end{table*}

\subsubsection{Environment A (CMLD Benchmark)}

Environment A consists of the IEEE 118-bus transmission system simulated in PSS/E, with a CMLD connected at bus~1. The ground-truth CMLD parameters are specified a priori and used to generate the disturbance measurements at 240~Hz. Only the resulting boundary measurements are provided to the parameter-estimation methods.

Because the estimation model has the same structure as the ground-truth load model, Environment A provides a controlled benchmark for evaluating both parameter recovery and dynamic-response accuracy. In addition to trajectory errors after reinsertion, the estimated parameters can be directly compared with their ground-truth values.

\subsubsection{Environment B (T\&D Co-Simulation)}

Environment B consists of the IEEE 118-bus transmission system simulated in PSS/E and a detailed IEEE 123-bus distribution feeder simulated in OpenDSS. The transmission and distribution systems are coupled through a stop-and-run co-simulation with data exchanged at 240~Hz \cite{cosim,paduani}.

The distribution system is scaled to approximately 55.92~MW to match the load connected at transmission bus~1, requiring approximately 16 parallel replicas of the IEEE 123-bus feeder. Each feeder contains 91 load nodes with heterogeneous load compositions. Every load node is itself a CMLD whose nominal composition and parameters equal the Environment-A ground-truth values (Table~\ref{tab:paramfull}). To introduce spatial diversity, the non-static load fractions at individual nodes are independently varied within $\pm30\%$ of their nominal composition. A fixed random seed is used for reproducibility. Each feeder is connected through a 6-MVA substation transformer with 8\% reactance.

Unlike Environment A, the ground truth in Environment B is the detailed distribution network rather than a predefined CMLD. Consequently, no unique ground-truth CMLD parameter vector is assumed to exist. This environment evaluates the ability of an estimated aggregate CMLD to reproduce the terminal dynamic behavior of a heterogeneous distribution system after reinsertion.

\subsection{Identification and Validation Protocol}

A single disturbance is used for parameter estimation in both environments. The identification event is a five-cycle, three-phase fault applied at bus~1 at $t=1$~s. The boundary-bus voltage and active and reactive powers are recorded at 240~Hz over a 15-s simulation window. All estimation methods use only the measurements obtained from this event.

For a consistent comparison, all methods use the same CMLD parameter space defined in \eqref{eq:theta}. The lower and upper bounds are set to 0 and 200\% of the nominal value, respectively, for parameters with nonzero nominal values. For parameters with a nominal value of zero, the search interval is set to $[-0.4,0.4]$. The same physical-validity constraints are imposed on all candidate parameter sets. As introduced in Section~II, the proposed framework adopts CMA-ES \cite{hansen} as the numerical solver for the CMLD parameter update in \eqref{eq:theta_update}. It uses the same settings as the conventional CMA-ES baseline, namely a population size of 20 (i.e., $N_P=20$) and an initial step size of $\sigma_0=0.25$ in the normalized parameter space (Table~\ref{tab:protocol}), so that all methods operate under a common computational budget.

Following parameter estimation, each identified CMLD is reinserted into the original simulation environment. In Environment A, reinsertion is performed in the full IEEE 118-bus system; in Environment B, it is performed in the original T\&D co-simulation. Reinsertion accuracy is first evaluated using the bus-1 identification event.

To assess whether the estimated model reproduces system dynamics beyond the disturbance used for identification, three additional three-phase faults are applied at buses~3, 4, and 5. These contingencies are not used during parameter estimation and are therefore treated as unseen validation events. For all four contingencies, the voltage, active-power, and reactive-power errors ($E_V^{\mathrm{reins}}$, $E_P^{\mathrm{reins}}$, and $E_Q^{\mathrm{reins}}$) are evaluated at bus~1, corresponding to the interface at which the aggregate CMLD is connected.

\subsection{Replay-to-Reinsertion Gap in the CMLD Benchmark}
\begin{figure*}[!t]
\centering
\includegraphics[width=\textwidth]{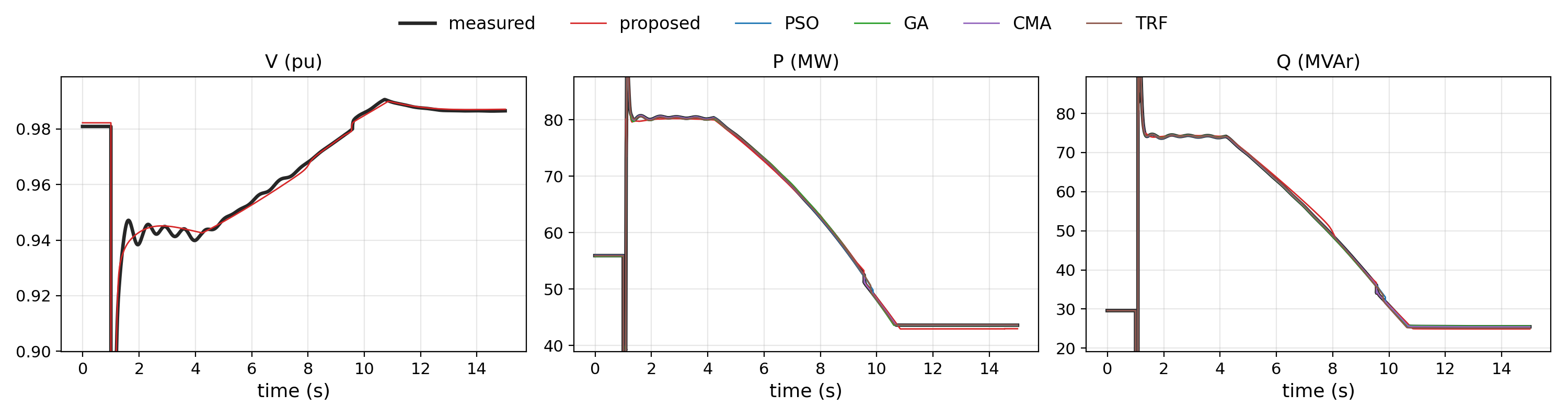}
\caption{Fitting results in Environment A at the identification contingency (signals at bus 1).}
\label{fig:fitA}
\vspace{2pt}
\includegraphics[width=\textwidth]{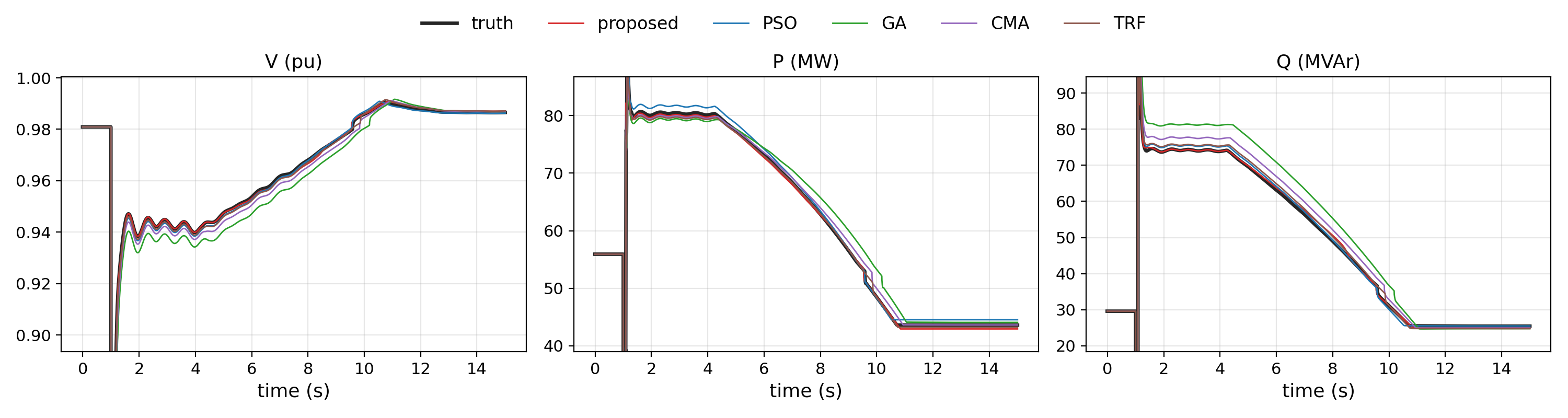}
\caption{Reinsertion time series in Environment A at the identification contingency (signals at bus 1).}
\label{fig:reinsA}
\end{figure*}

Environment A first examines whether a small voltage-replay error translates into an accurate CMLD after reinsertion. As shown in Fig.~\ref{fig:fitA}, all four conventional methods closely reproduce the measured $P$ and $Q$ trajectories when the measured voltage is replayed. In terms of the replay objective $J_1=E_P^{\mathrm{rep}}+E_Q^{\mathrm{rep}}$ in \eqref{eq:j1}, CMA-ES reaches approximately $10^{-7}$, and the remaining methods also drive $J_1$ to the $10^{-4}$ range; that is, the replay errors $E_P^{\mathrm{rep}}$ and $E_Q^{\mathrm{rep}}$ are driven nearly to zero. Nevertheless, once the corresponding parameter sets are reinserted into the IEEE 118-bus system, their reinsertion errors $E_P^{\mathrm{reins}}$, $E_Q^{\mathrm{reins}}$, and $E_V^{\mathrm{reins}}$ remain large and differ substantially across methods, as shown in Fig.~\ref{fig:reinsA} and Table~\ref{tab:reinsA}. This result demonstrates that replay accuracy alone is not a reliable indicator of reinsertion accuracy.

\begin{table*}[!b]
\centering
\caption{Reinsertion Comparison in Environment B: Errors at Bus 1 for Each Fault Bus ($E_P^{\mathrm{reins}}$ MW / $E_Q^{\mathrm{reins}}$ MVAr / $E_V^{\mathrm{reins}}$ p.u.)}
\label{tab:reinsB}
\footnotesize
\begin{tabular}{lcccc}
\toprule
\textbf{Method} & \textbf{Fault-at-Bus1 (Identification)} & \textbf{Fault-at-Bus3 (unseen event)} & \textbf{Fault-at-Bus4 (unseen event)} & \textbf{Fault-at-Bus5 (unseen event)} \\
\midrule
\rowcolor{lightgray}
Proposed & 0.61 / 1.13 / 0.0013 & 1.73 / 1.78 / 0.0017 & 1.54 / 2.25 / 0.0027 & 1.73 / 3.16 / 0.0034 \\
PSO & 3.19 / 3.15 / 0.0034 & 1.63 / 3.30 / 0.0031 & 1.91 / 2.24 / 0.0028 & 1.76 / 1.54 / 0.0023 \\
GA & 2.24 / 5.78 / 0.0048 & 1.88 / 5.89 / 0.0046 & 2.03 / 4.27 / 0.0037 & 2.38 / 3.29 / 0.0031 \\
CMA & 3.38 / 5.70 / 0.0055 & 2.78 / 6.13 / 0.0055 & 2.96 / 4.32 / 0.0045 & 3.13 / 4.10 / 0.0045 \\
TRF & 5.21 / 8.19 / 0.0079 & 3.78 / 8.39 / 0.0077 & 4.11 / 6.66 / 0.0067 & 3.96 / 5.89 / 0.0061 \\
\bottomrule
\end{tabular}
\end{table*}

The proposed closed-loop method provides more consistent accuracy after reinsertion. For the identification event at bus~1, its active-power error is comparable to the best conventional result, while it achieves the lowest reactive-power and voltage errors. The improvement is particularly evident in the post-fault reactive-power and voltage recovery shown in Fig.~\ref{fig:reinsA}. These results support the motivation for including the load-grid interaction during parameter estimation: candidate CMLD parameters are constrained not only by their $P$ and $Q$ responses but also by the voltage trajectory they produce through the grid interaction.

The same behavior is observed for the three unseen fault events at buses~3--5. The proposed method maintains similar $P$, $Q$, and $V$ errors across the identification and unseen contingencies, whereas the relative performance of the conventional replay-based estimates varies after reinsertion. The results therefore indicate that the proposed method improves not only the reproduction of the identification event but also the transferability of the estimated CMLD to disturbances not used during parameter estimation.

\subsection{Reinsertion Performance in T\&D Co-Simulation}

Environment B evaluates the methods under model mismatch, where the detailed distribution feeder serves as the ground truth and no exact CMLD parameter vector exists. Although the conventional replay-based methods again achieve small fitting errors, with replay objectives on the order of $10^{-3}$, their reinsertion performance differs substantially once the estimated CMLD is placed back into the T\&D co-simulation.

\begin{figure*}[!t]
\centering
\includegraphics[width=\textwidth]{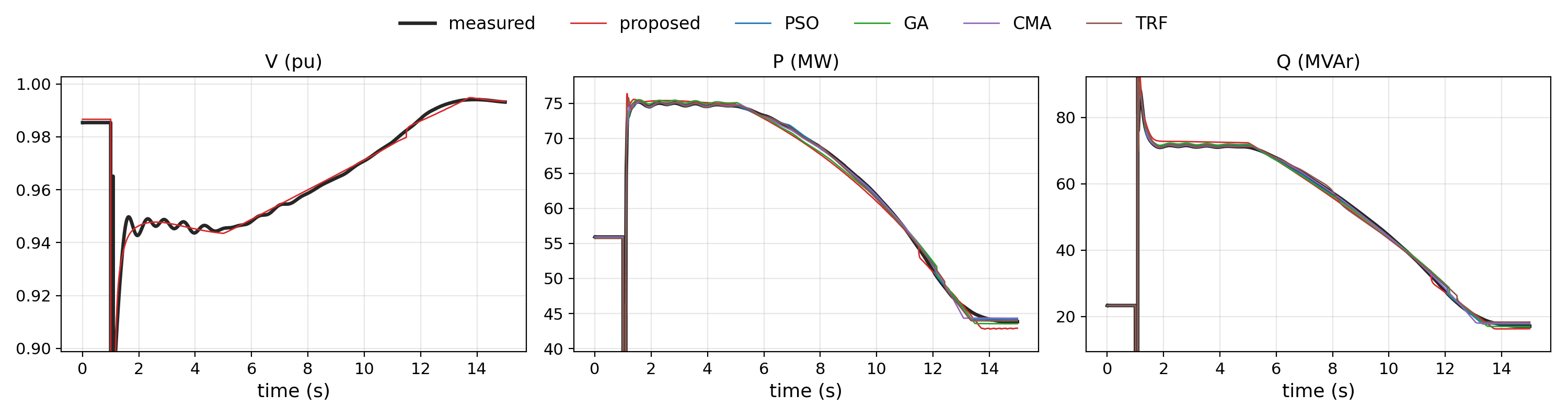}
\caption{Fitting results in Environment B at the identification contingency (signals at bus 1).}
\label{fig:fitB}
\vspace{2pt}
\includegraphics[width=\textwidth]{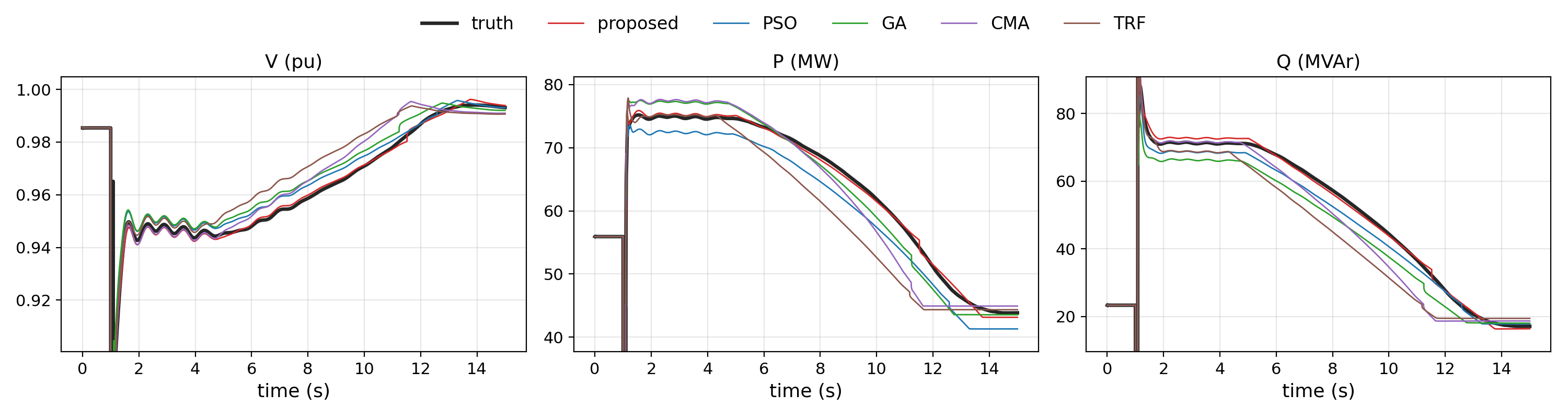}
\caption{Reinsertion time series in Environment B at the identification contingency (signals at bus 1).}
\label{fig:reinsB}
\end{figure*}

At the identification contingency, the proposed method achieves the lowest reinsertion errors in all three measured quantities, as summarized in Table~\ref{tab:reinsB}. Relative to the best conventional baseline for each metric, the $P$, $Q$, and $V$ errors are reduced by 73\%, 64\%, and 62\%, respectively. This improvement is consistent with the proposed formulation: because the candidate CMLD interacts with the equivalent network during estimation, the identified parameters are constrained by the coupled $P$--$Q$--$V$ response rather than by $P$ and $Q$ under a prescribed voltage trajectory alone.

The advantage also persists across the unseen contingencies. Averaged over the four fault locations, the proposed method yields $E_P^{\mathrm{reins}}=1.40$~MW, $E_Q^{\mathrm{reins}}=2.08$~MVAr, and $E_V^{\mathrm{reins}}=0.0023$~pu, compared with 2.12~MW, 2.56~MVAr, and 0.0029~pu for the best baseline averages. These correspond to reductions of approximately 34\%, 19\%, and 22\%, respectively. Although individual baselines may outperform the proposed method for a particular signal at a particular unseen contingency, the proposed method provides the lowest average error across all three quantities.

\begin{figure}[!t]
\centering
\includegraphics[width=\columnwidth]{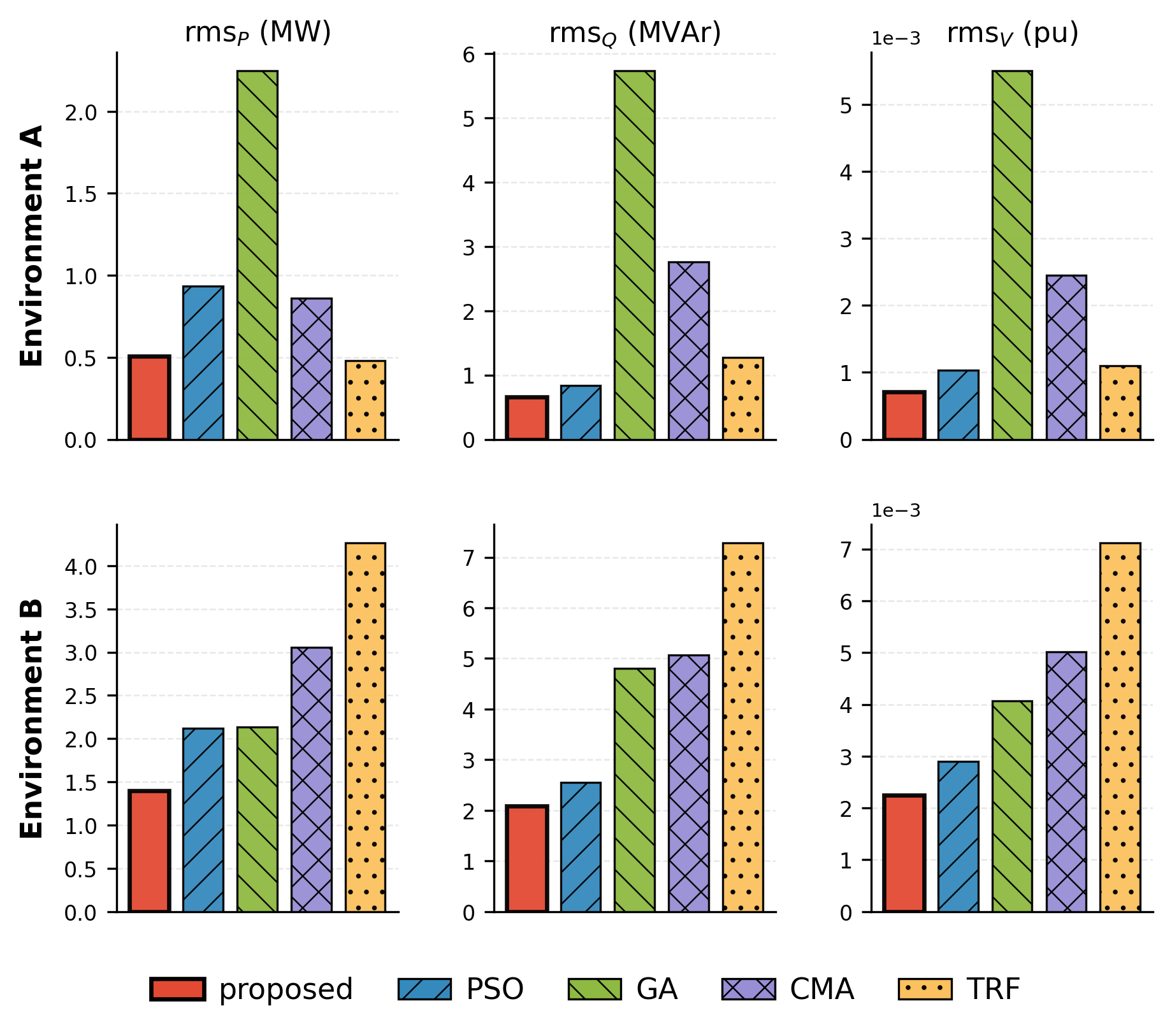}
\caption{Reinsertion errors in Environments A and B, averaged over the four contingencies (fault buses 1, 3, 4, 5).}
\label{fig:sumbars}
\end{figure}
\addtocounter{table}{1}
\begin{table*}[!b]
\centering
\caption{Per-Parameter Result of the Proposed Method in Environment A: True Versus Co-Estimated Values, Including the Equivalent-Network Parameters $\eta$}
\label{tab:paramfull}
\footnotesize
\setlength{\tabcolsep}{4.0pt}
\begin{tabular}{lr>{\columncolor[gray]{0.8}}rlr>{\columncolor[gray]{0.8}}rlr>{\columncolor[gray]{0.8}}rlr>{\columncolor[gray]{0.8}}r}
\toprule
\textbf{Param.} & \textbf{True} & \textbf{Est.} & \textbf{Param.} & \textbf{True} & \textbf{Est.} & \textbf{Param.} & \textbf{True} & \textbf{Est.} & \textbf{Param.} & \textbf{True} & \textbf{Est.} \\
\midrule
\multicolumn{3}{c}{\textbf{Load fractions}} & $V_{d2}$ & 0.5 & 0.2824 & $T_{ppo}$ & 0.0026 & 0.00195 & $N_{q1}$ & 2 & 3.737 \\
$F_{MA}$ & 0.264 & 0.4854 & \multicolumn{3}{c}{\textbf{Motor A}} & \multicolumn{3}{c}{\textbf{Motor C}} & $N_{q2}$ & 2.5 & 3.398 \\
$F_{MB}$ & 0.121 & 0.0302 & $X_s$ & 1.8 & 3.405 & $X_s$ & 1.8 & 1.568 & $K_{pf}$ & 1 & 0.6939 \\
$F_{MC}$ & 0.07 & 0.0204 & $X_p$ & 0.12 & 0.1854 & $X_p$ & 0.19 & 0.3408 & $K_{qf}$ & -3.3 & -1.165 \\
$F_{MD}$ & 0.198 & 0.1981 & $X_{pp}$ & 0.104 & 0.1228 & $X_{pp}$ & 0.14 & 0.2261 & $R_{stall}$ & 0.1 & 0.1064 \\
$F_{el}$ & 0.121 & 0.1715 & $R_a$ & 0.02 & 0.005241 & $R_a$ & 0.02 & 0.009704 & $X_{stall}$ & 0.1 & 0.1015 \\
\multicolumn{3}{c}{\textbf{Static load}} & $H$ & 0.1 & 0.09731 & $H$ & 0.1 & 0.2 & \multicolumn{3}{c}{\textbf{Feeder / tap}} \\
$P_{1c}$ & 0.112 & 0.0768 & $T_{po}$ & 0.08 & 0.1119 & $T_{po}$ & 0.2 & 0.3315 & $R_{fdr}$ & 0.04 & 0.04516 \\
$P_{2c}$ & 0.114 & 0.228 & $T_{ppo}$ & 0.0021 & 0.001575 & $T_{ppo}$ & 0.0026 & 0.003566 & $X_{fdr}$ & 0.04 & 0.03207 \\
$Q_{1c}$ & 1 & 0.4096 & \multicolumn{3}{c}{\textbf{Motor B}} & \multicolumn{3}{c}{\textbf{Motor D}} & $TAP$ & 1 & 1.02 \\
$Q_{2c}$ & 0 & 0.3614 & $X_s$ & 1.8 & 2.141 & $K_{p1}$ & 0 & 0.1841 & \multicolumn{3}{c}{\textbf{Equivalent network $\eta$}} \\
$P_{freq}$ & 0 & -0.3547 & $X_p$ & 0.19 & 0.2791 & $K_{p2}$ & 12 & 20.58 & $MBASE$ & -- & 1188.9 \\
$Q_{freq}$ & -1 & -0.6816 & $X_{pp}$ & 0.14 & 0.2527 & $K_{q1}$ & 6 & 2.743 & $K_{avr}$ & -- & 57.58 \\
\multicolumn{3}{c}{\textbf{Electronic load}} & $R_a$ & 0.02 & 0.01004 & $K_{q2}$ & 11 & 8.37 & $Z_r$ & -- & 0.004187 \\
$F_{rcel}$ & 0.8 & 0.6258 & $H$ & 0.5 & 0.5842 & $N_{p1}$ & 1 & 1.455 & $Z_x$ & -- & 0.0758 \\
$V_{d1}$ & 0.7 & 0.3765 & $T_{po}$ & 0.2 & 0.3129 & $N_{p2}$ & 3.2 & 1.812 & $V_{set}$ & -- & 0.9787 \\
\bottomrule
\end{tabular}
\end{table*}
\addtocounter{table}{-2}

The contrast between Figs.~\ref{fig:fitB} and~\ref{fig:reinsB} further illustrates the replay-to-reinsertion gap under model mismatch. During voltage replay, the conventional methods produce closely matched $P$ and $Q$ trajectories because all are driven by the same measured voltage. After reinsertion, however, their closed-loop responses separate substantially. In contrast, the proposed method reproduces the post-fault $P$, $Q$, and $V$ recovery more consistently because the load-grid interaction is already represented during parameter estimation.

The T\&D co-simulation results are particularly important because Environment B does not assume that the detailed feeder can be represented by one unique CMLD parameter vector. The objective is therefore not exact parameter recovery, but identification of an aggregate CMLD that preserves the dynamic behavior observed at the transmission-distribution interface. The improved reinsertion accuracy indicates that the proposed closed-loop formulation is better aligned with this system-level modeling objective than conventional voltage replay. Figure~\ref{fig:sumbars} summarizes the average reinsertion errors over the four contingencies for both validation environments.

\subsection{Parameter Accuracy Analysis}

Direct parameter-accuracy evaluation is possible only in Environment A because its ground-truth load is itself a CMLD with a known parameter vector. Environment B does not admit an exact ground-truth CMLD parameter set: the detailed distribution feeder contains spatially heterogeneous loads operating at different local voltages, and its aggregation into a single CMLD is inherently approximate. Therefore, parameter error is evaluated only for Environment A, whereas estimation quality in Environment B is assessed through reinsertion response accuracy.

For Environment A, the span-normalized mean absolute error (NMAE) in \eqref{eq:error} is decomposed by CMLD component in Table~\ref{tab:paramerr}. Parameters associated with load fractions, feeder impedance, and tap settings are included only in the overall NMAE because they are not uniquely associated with an individual component. Table~\ref{tab:paramfull} further reports the ground-truth and estimated values of the 50 CMLD parameters for the proposed method. The co-estimated equivalent-network parameters are also listed, although no ground-truth values are defined for these auxiliary parameters.

\begin{figure}[!b]
\centering
{\makeatletter\def\@captype{table}\makeatother
\caption{Component-Wise Parameter Error NMAE (\%; Environment A only, the environment with a ground-truth parameter vector)}
\label{tab:paramerr}
\footnotesize
\setlength{\tabcolsep}{3.5pt}
\begin{tabular}{lccccccc}
\toprule
\textbf{Method} & \textbf{Mot.A} & \textbf{Mot.B} &\textbf{ Mot.C} & \textbf{Mot.D} & \textbf{PE} & \textbf{ZIP} & \textbf{Overall} \\
\midrule
\rowcolor{lightgray}
Proposed & 15.8 & 19.2 & 21.7 & 27.2 & 11.9 & 38.3 & 21.3 \\
PSO & 15.9 & 24.4 & 15.1 & 19.4 & 14.0 & 21.1 & 19.3 \\
GA & 10.4 & 6.4 & 19.6 & 25.8 & 23.1 & 37.6 & 19.8 \\
CMA & 9.1 & 20.0 & 22.5 & 20.8 & 3.4 & 30.0 & 17.4 \\
TRF & 17.9 & 10.1 & 18.9 & 23.9 & 16.3 & 25.8 & 18.2 \\
\bottomrule
\end{tabular}}
\end{figure}
\addtocounter{table}{1}

The results indicate that accurate recovery of individual CMLD parameters remains difficult even when the model structure is known exactly. No method is uniformly superior across all components, and the component-wise ranking varies substantially among the estimators. More importantly, the overall parameter error is not strongly correlated with reinsertion accuracy. The overall NMAE of the tested methods falls within a relatively narrow range of approximately 17--21\%, yet their reinsertion performance differs substantially. For example, the proposed method achieves the best overall reinsertion performance despite having an overall parameter NMAE of 21.3\%, whereas CMA-ES attains the lowest parameter NMAE of 17.4\% but produces larger reinsertion errors.

This result provides further evidence of parameter non-uniqueness in measurement-based CMLD estimation. Different parameter sets can reproduce similar disturbance trajectories under voltage replay while differing considerably from the ground-truth parameters and from one another. Consequently, parameter recovery and dynamic-response reproduction should be treated as distinct evaluation objectives. In practical applications such as Environment B, where no unique ground-truth CMLD parameter vector exists, reinsertion accuracy becomes the more meaningful criterion for assessing the quality of the identified aggregate load model.

\subsection{Computational Cost Analysis}

The closed-loop formulation introduces additional computational cost because each candidate CMLD parameter set must be evaluated through a dynamic simulation of the equivalent network rather than through open-loop voltage replay. In Environment A, one closed-loop evaluation requires approximately 4.6~s, compared with approximately 0.2~s for one conventional replay evaluation. This increase represents the principal computational overhead of the proposed method.

Table~\ref{tab:cost} compares the total wall-clock times of the proposed and baseline methods under the computational settings used in this study. The proposed method requires approximately 11.9~h, compared with 2.0--6.1~h for the conventional replay-based methods. Thus, the improved reinsertion accuracy of the proposed method is obtained at the expense of increased offline parameter-estimation time. The additional cost arises primarily from restoring the load-grid interaction during each candidate evaluation rather than from the parameter-search algorithm itself.

Because CMLD parameter identification is performed offline and the proposed method uses only a low-order equivalent network rather than the full transmission system during optimization, the resulting computational requirement remains practical for the intended application. Further reductions may be obtained through parallel evaluation of candidate parameter sets and more efficient implementation of the closed-loop simulator.

\begin{center}
\begin{minipage}{\columnwidth}
\makeatletter\def\@captype{table}\makeatother
\centering
\caption{Wall-Clock Time for CMLD Parameter Estimation in Environment A}
\label{tab:cost}
\scriptsize
\setlength{\tabcolsep}{2.5pt}
\begin{tabular}{l>{\columncolor[gray]{0.8}}ccccc}
\toprule
\textbf{Item} & \textbf{Proposed Method} & \textbf{PSO} & \textbf{GA} & \textbf{CMA} & \textbf{TRF} \\
\midrule
Time (s) & $4.3\times10^4$ & $2.2\times10^4$ & $7.1\times10^3$ & $8.6\times10^3$ & $1.0\times10^4$ \\
\bottomrule
\end{tabular}
\end{minipage}
\end{center}

\section{Conclusion}

This paper demonstrated that a small voltage-replay error does not necessarily lead to an accurate CMLD after reinsertion into dynamic grid simulations. To address this replay-to-reinsertion gap, a closed-loop co-estimation framework was proposed in which a low-order equivalent network restores the load-grid interaction and enables $P$, $Q$, and $V$ to jointly constrain the estimated CMLD. Validation using a controlled CMLD benchmark and a detailed T\&D co-simulation showed improved reinsertion accuracy over conventional replay-based methods. In the T\&D co-simulation, the proposed method reduced the $P$, $Q$, and $V$ errors by 73\%, 64\%, and 62\%, respectively, for the identification event, while maintaining improved average performance across unseen contingencies. These results indicate that reinsertion accuracy provides a more meaningful measure of aggregate load-model quality than replay error alone.

Future work will investigate the sensitivity of the proposed framework to the structure and parameterization of the equivalent network, including alternative equivalent-source models and reduced-order representations. The method will also be evaluated using field PMU measurements to assess its performance under practical measurement uncertainty, unmodeled load dynamics, and real-world disturbances.

\bibliographystyle{IEEEtran}
\bibliography{references}

\end{document}